\documentclass{emulateapj}

\newcommand{\kpc}{~h^{-1}~kpc}
\newcommand{\Mpc}{~h^{-1}~Mpc}
\newcommand{\Msun}{~h^{-1}~M_{\odot}}

\slugcomment{}

\shorttitle{Environmental Dependence of Galaxy Merger Rate}

\shortauthors{ Jian, Lin \& Chiueh}

\begin{document}

\title{Environmental Dependence of Galaxy Merger Rate in $\Lambda CDM$ Universe}

\author{Hung-Yu Jian \altaffilmark{1}, Lihwai Lin\altaffilmark{2}, and Tzihong Chiueh\altaffilmark{1,3,4}}
%\email{hyj@phys.ntu.edu.tw, lihwailin@asiaa.sinica.edu.tw \& chiuehth@phys.ntu.edu.tw}
%\affil{$^{1}$Department of Physics, National Taiwan University, 106, Taipei, Taiwan, %R.O.C.}
%\affil{$^{2}$Institute of Astronomy Astrophysics, Academia Sinica, 106, Taipei, Taiwan, %R.O.C.}
%\affil{$^{3}$Center for Theoretical Sciences, National Taiwan University, 106, Taipei, %Taiwan, R.O.C.}
%\affil{$^{4}$LeCosPa, National Taiwan University, 106, Taipei, Taiwan, R.O.C.}
%\email{hyj@phys.ntu.edu.tw, lihwailin@asiaa.sinica.edu.tw \& chiuehth@phys.ntu.edu.tw}

\altaffiltext{1}{Department of Physics, National Taiwan University, 106, Taipei, Taiwan, R.O.C.}
\altaffiltext{2}{Institute of Astronomy Astrophysics, Academia Sinica, 106, Taipei, Taiwan, R.O.C.}
\altaffiltext{3}{Center for Theoretical Sciences, National Taiwan University, 106, Taipei, Taiwan, R.O.C. }
\altaffiltext{4}{LeCosPa, National Taiwan University, 106, Taipei, Taiwan, R.O.C.}

\begin{abstract}
We make use of four galaxy catalogs based on four different semi-analytical models (SAMs) implemented in the Millennium simulation to study the environmental effects and the model dependence of galaxy merger rate. We begin the analyses by finding that galaxy merger rate in the SAMs has mild redshift evolution with luminosity selected samples in the evolution-corrected B band magnitude range,$-21 \leq M^{e}_{B} \leq -19$, consistent with results of previous works. To study the environmental dependence of galaxy merger rate, we adopt two estimators, the local overdensity $(1+\delta_{n})$ defined as the surface density from the $n^{th}$-nearest-neighbor ($n = 6$ is chosen in this study) and the host halo mass $M_{h}$. We find that galaxy merger rate $F_{mg}$ shows strong dependence on the local overdensity $(1+\delta_n)$ and the dependence is similar at all redshifts. For the overdensity estimator the merger rate $F_{mg}$ is found about twenty times larger in the densest regions than in under-dense ones in two of the four SAMs while it is roughly four times higher in the other two. In other words, the discrepancies of the merger rate difference between two extremes can differ by a factor of $\sim$ five depending on the SAMs adopted. On the other hand for the halo mass estimator, $F_{mg}$ does not monotonically increase with the host halo mass $M_h$, but peaks in the $M_h$ range between $10^{12}$ and $10^{13}$ $\Msun$, which corresponds to group environments. High merger rate in high local density regions corresponds primarily to the high merge rate in group environments. In addition, we also study the merger probability of "close pairs" identified using projected separation and line-of-sight velocity difference $C_{mg}$ and the merger timescale $T_{mg}$; these are two important quantities for observations to convert the pair fraction $N_c$ into galaxy merger rate. We discover that $T_{mg}$ has weak dependence on environment and different SAMS, and is about 2 $Gyr$ at $z$ $\sim$ 1. In contrast, $C_{mg}$ depends both on environment (declining with density) and different SAMs; its environmental dependence is primarily due to the projection effect. At $z~\sim~1$, it is found that only $\sim~31\%$ of projected close pairs will eventually merge by $z~=~0$. We find that the projection effect is the dominant factor to account for the low merger probability of the projected close pairs.    \end{abstract}


\section{Introduction}
In a standard $\Lambda$ cold dark matter ($\Lambda CDM$) model, the major framework is the hierarchical structure formation that dark matter halos and galaxies within them are assembled from the successive accretion of smaller objects \citep{pee82,blu84,dav85}. Merger is thus an inevitable process and plays an essential role in galaxy formation and evolution. The potential impacts of mergers on galaxy evolution have been intensely discussed. In additional to being responsible for mass growth in the assembly history of galaxies, they are, for example, expected to trigger quasar activities \citep{car90} and starbursts \citep{bar91}, and thought to shape many important observational properties of galaxies, such as star formation rate, color and morphology transformation \citep{too72}. Galaxy mergers continue to be an important component for galaxy evolution from more recent works on star formation history and remnant properties \citep[e.g.][]{cox06}, starbursts, quasars, and black hole growth \citep[e.g.][]{hop06}, bulge growth and morphology \citep[e.g.][]{hop10a},  baryonic content of galaxy \citep[e.g.][]{ste09a}, and formation of diffuse intra-halo light components \citep[e.g.][]{pur07}.

Many studies have explored the galaxy merger rate as a function of cosmic time to understand the assembly history of galaxies observationally \citep{pat02,con03,lin04,lin08,lotz08,lotz11,kam07,kar07,hsieh08,bun09,deRav09,deRav11,chou11,xu11} and theoretically \citep[e.g.,][]{ber06,mate08,ste09,hop10}. Since galaxy merger assembles two or more galaxies into one, it is expected to be more common in dense regions, i.e., having strong dependence on environments. To what degree this can be so is an issue of great concern. The observational work of \cite{lin10} was the first study to address this issue. They probed the environments of wet, dry, and mixed galaxy mergers at 0.75 $<$ $z$ $<$ 1.2 using close pairs in the DEEP2 Galaxy Redshift Survey, and found that the environment dependence of the galaxy merger rate for wet mergers is marginal while the dry and mixed merger rates increase rapidly with local density. A similar conclusion that merger rate has strong environment dependence latter was also obtained by \cite{deRav11} and \cite{kam11}. However, galaxy merger rate in observation is actually an indirect physical quantity and can only be obtained by combining the fraction of galaxies that reveal signatures of merger activity from observations, and the merger time scale and the merger probability estimated from theoretical models. The environmental dependence of the merger probability and the merger timescale in theoretical models is less addressed especially. Hence, to interpret the observational results more precisely, theoretical determination on how the merger time varies with environments and on how much pair contamination is due to projection effect in terms of environment is essential. There have been only several theoretical studies to address environmental dependence associated with mergers \citep{fak09,hes10,ton11}. \cite{fak09} utilized the Millennium Simulation to derive the merger rate of FOF DM halos as a function of the local mass density within a sphere of 7 $\Mpc$, and demonstrated that mergers occur about 2.5 times more frequently in the densest regions than in voids at both $z$ = 0 and higher redshifts. However, their study focused on mergers of distinct FOF halos, while galaxy mergers do not necessarily follow their scenario. \cite{hes10} alternatively adopted the merger tree of subhalos from the Millennium database to explore the same issues. The local halo environment in their study is defined as the count of all distinct halos and sub-halos, with $V_{max}~\leq $ 120 $km s^{-1}$ within a sphere of comoving radius 2 $\Mpc$ centered on each halo. They found that the merger rate is roughly independent of environment. However, the evolutionary history of subhalos may not exactly be that of the galaxies, due to different treatments on on orphan galaxies when subhalos are disrupted which may lead to different conclusions. The requirements for a relation between the histories of a given dark matter halo and of a galaxy inside that halo make the conversion from dark matter (halo/subhalo) merger rates to galaxy merger rates non-trivial \citep{hop10}.

In this study, we make use the publicly available galaxy catalogs based on four different semi-analytical models (SAMs) implemented in the Millennium database. In addition to exploring the environmental dependence of galaxy merger rate, we also investigate the merger probability of close pairs $(C_{mg})$, which provides a correction due to the projected selection criterion, and the merger timescale $(T_{mg})$, which provides the mean merger time for a close pair to merge, as a function of environments. These two quantities are important for observations to convert the pair fraction into galaxy merger rate. Besides, it is illuminating to understand how different SAMs affect galaxy merger rate and these two quantities. In fact, \cite{hop10} already pointed out that different theoretical methodologies lead to order-of-magnitude variations in predicted galaxy-galaxy merger rates. Whether such variations among different SAMs propagate into the environment dependence requires careful scrutiny.

The paper is structured as follows. We briefly describe the Millennium simulation and the four galaxy catalogs from the semi-analytical models in use in Section 2. Section 3 is divided into five sub-sections. Environmental estimators are introduced in Section 3.1. In Section 3.2, galaxy merger rate is defined, where the redshift evolution of the pair fraction $N_{c}$ and galaxy merger rate between the models and observations are compared. The environmental dependence of $N_{c}$ and galaxy merger rate is presented in Section 3.3. Our results on $C_{mg}$, and $T_{mg}$ as a function of environment are shown in Section 3.4. In Section 3.5, the impact of mass ratio of merging galaxies on the merger rate in  different environments is discussed. Finally, discussion and conclusions are given in Section 4 and 5, respectively.

\section{Models}

\subsection{Simulations} \label{Simu}
We make use of the Millennium Simulation, provided by the Virgo Consortium, in this study. The Millennium Simulation follows the hierarchical growth of dark matter (DM) structures with $N = 2160^{3}$ particles of mass $8.6 \times 10^{8} \Msun$ in a co-moving periodic box of size $500  \Mpc$ on a side from redshift $z$ = 127 to the present \citep{spr05} where the Hubble constant is parameterized as $H_{0} = 100~h~km~s^{-1}~Mpc^{-1}$. The concordance $\Lambda CDM$ cosmology is assumed in the Millennium simulation with parameters $\Omega_{m} = 0.25$, $\Omega_{b} = 0.045$, $h = 0.73$, $\Omega_{\Lambda} = 0.75$, $n = 1$, and $\sigma_{8} = 0.9$, consistent with a combined analysis of the 2dF Galaxy Redshift Survey and the first-year Wilkinson Microwave Anisotropy Probe (WMAP) data. The GADGET-2 code \citep{sprin05}, a variant of the TreePM method, is adopted for evaluating gravitational forces, allowing for a very large dynamic range and high computational speed even in situations where the clustering becomes strong. With a soften length of $5 \kpc$ on a comoving scale, taken as the spatial resolution limit of the computation, the effective dynamic range thus reaches $10^5$ in spatial scale. In addition, haloes hosting all luminous galaxies brighter than $0.1~L_{*}$ (where $L_{*}$ is the characteristic luminosity of galaxies) can be resolved with a mass resolution of $\sim 100$ particles. There are 64 epochs of output data spaced approximately logarithmically in time at early epoches and approximately linearly in time at late epoches (with $\Delta~t$ $\sim$ 300 $Myr$).

A friends-of-friends (FOF) group finder \citep{dav85} with a linking length of b = 0.2 in units of the mean particle separation is used to identify halos in the simulation. Each FOF identified halo is further broken into constituent subhalos (each with at least 20 particles or $2.35 \times 10^{10} \Msun$) by the SUBFIND algorithm \citep{spr01} which identifies gravitationally bound substructures within the host FOF halo; SUBFIND typically finds a large background host subhalo and a number of smaller satellite subhalos. With all halos and subhalos determined, the hierarchical merging trees containing the details of how structures build up over cosmic time can then be constructed. These trees are the key information needed to compute the physical properties of the associated galaxy population for SAMs. It is known that the merger tree construction basically is to connect subhalos across all snapshot outputs, and established through the determination of a unique descendant from any given halo. In the Millennium simulation, each potential descendant is scored based on a weighted count giving higher weight to particles that are more tightly bound in the halo under consideration, and the unique descendant is the one with the highest score. In this way, the merger trees are defined and constitute the basic input needed by SAMs.

The other simulation, $\Lambda CDM_{100b}$, adopted in this work is evolved in the flat $\Lambda$CDM model: $\Omega_{0}=0.3, \Omega_{\Lambda}$ = 0.7, $\Omega_{b} = 0.05$, $h = 0.7$, and $\sigma_{8}=0.94$, using $512^{3}$ DM particles in a co-moving periodic box of size $100  \Mpc$ on a side from redshift z = 100 to the present \citep{jian08}. This simulation also employs the GADGET-2 code, and the softening length $\epsilon$ is set to $10\kpc$ (comoving) before redshift $z = 2.3$ and modified to $3\kpc$ (physical) thereafter. Additionally, the outputs are at 67 epochs spaced equally in 100 $\Mpc$ comoving length, and therefore, the time difference for two adjacent snapshots increases slightly. For example, $\Delta~t$ $\sim$ 0.24 $Gyr$ at $z$ $\sim$ 1 and $\Delta~t$ $\sim$ 0.4 $Gyr$ at $z$ $\sim$ 0.

In the simulation $\Lambda CDM_{100b}$, a subhalo finder called HiFOF \citep [Hierarchical Friends-of-Friends;][]{jian08} is hired to identify bound substructures. HiFOF first finds halos with a linking length of b = 0.2 in units of the mean particle separation (the lowest level), and starts to shorten the linking length linearly like a zoom-in process and perform virial condition check on those newly found substructures satisfying a criterion that the minimum particle number for subhalos has to be greater than 30, corresponding to $1.9\times 10^{10} \Msun$. The process will continue for both those bound or unbound structures until it reaches a certain level or find no further bound structures. The merger tree in this simulation is defined differently from the Millennium one in how the unique descendant is determined. The criterion is that potential descendants are the subhalos at least with 30 percent members overlapped with their progenitor members, and the descendant with most particles is the unique descendant. We also test the same tree definition as the one in Millennium and the result turns out to be similar. For example, the merger rate derived from these two trees differs less than $4\%$ for any redshift. Moreover, the galaxy model applied to the simulation $\Lambda CDM_{100b}$ is a simplified one \citep[see][for detail]{jian08} that subhalos residing in the host halos more massive than a certain mass $\sim$ $4\times 10^{11} \Msun$ at $z = 0$ are galaxy samples, and others are dark halos. This threshold is not based on galaxy luminosity but on the mass of the host halo in which galaxies reside. Due to the simple treatment for the galaxy properties, no orphan galaxies are considered, and the result from this simulation included in this analysis is to provide an extreme reference case relative to SAMs. The result allows us to understand whether the merger rate derived from subhalo mergers (without orphan galaxies) may be significantly different from the merger rate derived from galaxy mergers.

\subsection{Semi-analytic models} \label{SAMs}
Semi-analytic modeling on galaxy formation and evolution was initially proposed by \cite{whi01}, and since then variations of the original SAM have been proposed and verified to be a successful \citep{crot06,bow06,delu07,bert07,font08}. In this study, we focus on the galaxy catalogs produced by four different SAMs implemented on the Millennium simulation including Bower06 \citep{bow06}, Font08 \citep{font08}, DeLucia06 \citep{delu07}, and Bertone07 \citep{bert07}. We investigate how much the results converge for the galaxy merger rate as a function of environment among different models. In addition, these galaxy catalogs are utilized to obtain information such as the merger probability of close pairs and the merger time which are essential for observations to determine the galaxy merger rate. These catalogs are publicly available on the Millennium download site\footnote{http://galaxy-catalogue.dur.ac.uk:8080/Millennium/}. In the following, brief descriptions on different SAMs are given. For detail, the readers are refereed to the these papers.

Bower06 is based on the Durham semi-analytic model of galaxy formation, GALFORM, and it implements a new treatment for the active galactic nuclei (AGN) energy injection, determined by a self-regulating feedback loop that is assumed to quench cooling flows in massive haloes. They found that the feedback mechanism naturally produces a break at the bright end of the local luminosity function. Besides, satellite galaxies are assumed to lose all of their hot gases to the new parent halo upon being sufficiently close to the parent halos.

DeLucia06 is a slightly modified version of the one used in \cite{spr05}, \cite{crot06}, and \cite{delu07}. It includes the AGN feedback model from \cite{crot06} to suppress the cooling flows, and the model seems to be extremely efficient in switching off cooling in relatively massive haloes even at high redshifts. Additionally, they also employed the supernova feedback model from \cite{crot06} to drive strong winds which blow out all gases away from galaxies on short time-scales.

Bertone07 presents a new feedback scheme, which replaces empirical prescriptions on SN feedback of the Munich semi-analytic model with a dynamical treatment of the galactic wind evolution, where ejection and recycling of gases and metals are treated self-consistently. It was shown that the observed mass-stellar metallicity and luminosity-gas metallicity relationships are able to be reproduced by their model. However, two drawbacks exist in their model. The number of bright galaxies in the LFs tends to be overestimated, and the color distribution of galaxies does not display the sharp color bimodality observed for galaxies in the local universe.

Finally, in Font08 which is one of the Durham SAMs, the major change is that they merge a model of stripping processes based on detailed hydrodynamic simulations into GALFORM semi-analytical model for galaxy formation, in an attempt to fix the problem that satellite galaxies in groups and clusters are redder than observed. They found that the effect of the ram-pressure stripping on the colors of satellite galaxies is significant. With assumptions of the gradual stripping as opposed to the sudden stripping in Bower06, a considerable fraction of hot gases in satellite galaxies is preserve for several $Gyr$, and satellite galaxies remain blue for a relatively long period of time as a result.

\cite{kit08} noticed that different treatments in SAM influence the merger rate indirectly when identifying the merging systems as the observed merging galaxies. The galaxy formation modeling does not alter the dynamics of the underlying distribution of DM halos and subhalos, but when a galactic subhalo is tidally disrupted near the center of a more massive halo and eligible to merge into that halo, the model estimates the merger time based on a dynamical friction time argument rather than assuming instant merging. In other words, once the subhalo disrupts, the galaxy evolution model waits for one dynamical friction time before the associated galaxy of this halo completes merging into the central galaxy of the main halo. After subhalo disruption, the associated satellite galaxy is assumed to be attached to some particles, which are the most strongly bound particles of its last identified subhalo. The condition to obtain realistic close pairs is demonstrated in \cite{kit08} via comparisons on the two-point correlations between simulated galaxies and real galaxies. However, we notice that different treatments of orphan galaxies affect not only the time scales when a close pair completes merging but also the fraction of close pairs that will actually merge (see Section~\ref{CmgTmg}).

\section{Galaxy merger rate as a function of environment}

\begin{figure}
 \begin{center}
  \includegraphics[width=9cm]{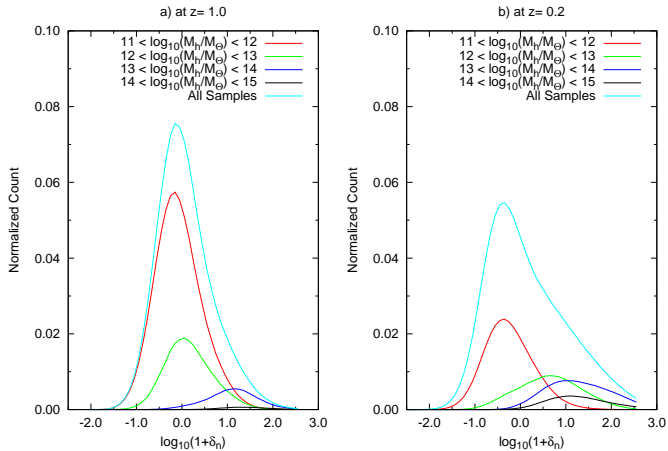}\\
  \caption{The distributions of the local density $(1+\delta_{n})$ are plotted in different $M_{h}$ (the host halo mass) bins represented by color lines at two redshifts (a) $z$ = 1.0 and (b) $z$ = 0.2. It is observed that for a given overdensity range, galaxies are from various host environments, such as filed, group, or cluster environment. At high density region, the main contribution is from galaxies in group or cluster environment. On the contrary, filed galaxies contribute significant amount at low density area. It is also seen that the relative contributions from galaxies in different environment evolve with redshift. Two environmental estimators are found to be well correlated.
   }\label{fig:MhDelta3}
 \end{center}
\end{figure}

\subsection{Environmental estimators: $M_{h}$ and $(1+\delta_{n})$} \label{Mhdeltan}
To study the environmental effect, we adopt two different environmental estimators. One is the host halo mass $M_{h}$ for the merging galaxies, which is a clear physical variable to represent the environmental effect and can be obtained easily from the simulation. However, the host halo mass is not a direct observable, and hence an environmental observable related to the local overdensity density $(1+\delta_{n})$ is also studied in conjunction with $M_{h}$. The local galaxy density $(1+\delta_{n})$ is estimated using the projected $n^{th}$-nearest-neighbor surface density, $\Sigma_{n} \equiv n/\pi D^{2}_{p,n}$, where $D_{p,n}$ is the projected distance to the $n^{th}$-nearest neighbor that is within the line-of-sight velocity interval of $¡Ó1000~km~s^{-1}$. In addition, each density value is divided by the median $\Sigma_{n}$ of galaxies at that redshift over the whole sample, i.e., $1 + \delta_{n} \equiv \Sigma_{n}/ median(\Sigma_{n})$ to reduce the influence of variations in redshift sampling in the survey \citep{coo05}. Note that, $(1+\delta_{n})$ is a relative but not a absolute quantity, and direct comparison for the same $(1+\delta_{n})$ at two different redshifts is, thus, not appropriate. In order to make direct comparison with DEEP2 results from \cite{lin10}, $n = 6$ is chosen to account for using the $3^{rd}$-nearest neighbor in the DEEP2 sample with the average redshift completeness $\sim 50\%$. In the rest of this paper, we refer 1+$\delta_{n}$ to the overdensity measured up to the $6^{th}$-nearest-neighbor for galaxies. Furthermore, the applied galaxy selection cuts are different in the simulation $\Lambda CDM_{100b}$ and in the SAMs. For the SAMs, galaxy samples are selected with a flux limit cut $R \leq 24.1$ which is the same cut as in DEEP galaxies. For simulation $\Lambda CDM_{100b}$, the median distance to the $n^{th}$-nearest-neighbor above certain mass cut $M_{cut}$ for subhalos is computed so as to match the median distance to the $3^{rd}$-nearest neighbor in the DEEP2 sample. The resulting $(1 + \delta_{n})$ distribution of halos for $\Lambda CDM_{100b}$ displays a very similar profile with the $(1 + \delta_{3})$ distribution of observed DEEP2 galaxies \citep[see Figure 4 of ][]{lin10}. The corresponding $M_{cut}$ for $n = 6$ are (a) 2.23, (b) 2.48, (c) 5.75, and (d) 8.35 $\times ~10^{10} \Msun$ for four different redshifts, i.e. z = (a) $\sim$ 1.1, (b) $\sim$ 0.9, (c) $\sim$ 0.7, and (d)$\sim$ 0.5.

With these two estimators $M_{h}$ and $(1+\delta_n)$, it is interesting to understand how they are correlated and how their correlation evolves. Therefore, the local density $(1+\delta_n)$ distributions for galaxies resided in hosts of different masses denoted in color-coded lines from Font08 are plotted in Figure~\ref{fig:MhDelta3} as an example. We split a total sample of N galaxies at a certain redshift into four bins based on the host halo mass of galaxies. These $M_{h}$ bins are $11 \leqslant log_{10}(M_{h}/(\Msun)) \leqslant 12$ (red), $12 \leqslant log_{10}(M_{h}/(\Msun)) \leqslant 13$ (green), $13 \leqslant log_{10}(M_{h}/(\Msun)) \leqslant 14$ (blue), and $14 \leqslant log_{10}(M_{h}/(\Msun)) \leqslant 15$ (black). In each of these four $M_h$ bins, we further divide galaxies into different $(1+\delta_n)$ bins. That is, the total sum of galaxy counts from all $(1+\delta_n)$ bin and from four $M_h$ bins should equal to N. We subsequently use N as a normalization factor so that the sum of normalized count from all $(1+\delta_n)$ bins for the light blue line is equal to 1. It is noted that for a given range of local density $(1+\delta_n)$, its total strength is a collective result combing different contributions from various hosting environments. In other words, a galaxies with the highest $(1+\delta_n)$ is not necessary from the most massive hosting environment. Nevertheless, it is seen in Figure~\ref{fig:MhDelta3} that for those galaxies in more massive host halo, their local density $(1+\delta_n)$ tends to be distributed in a denser regions. In addition, it is observed that when comparing the local density distributions at two different redshifts that the relative contribution from different $M_h$ bins evolves with redshift. The contribution from the two most massive $M_h$ bins increases with $z$ and that from the two least massive bins decreases with $z$, implying that in time field galaxies gradually merge with galaxy clusters. Therefore, we conclude that two estimators are well-correlated and the local density indicator $(1+\delta_n)$ has a real physical meaning for the environment, and the dominant contribution for large $(1+\delta_n)$ is from groups or small clusters.

\begin{figure*}
 \begin{center}
  \includegraphics[width=15cm]{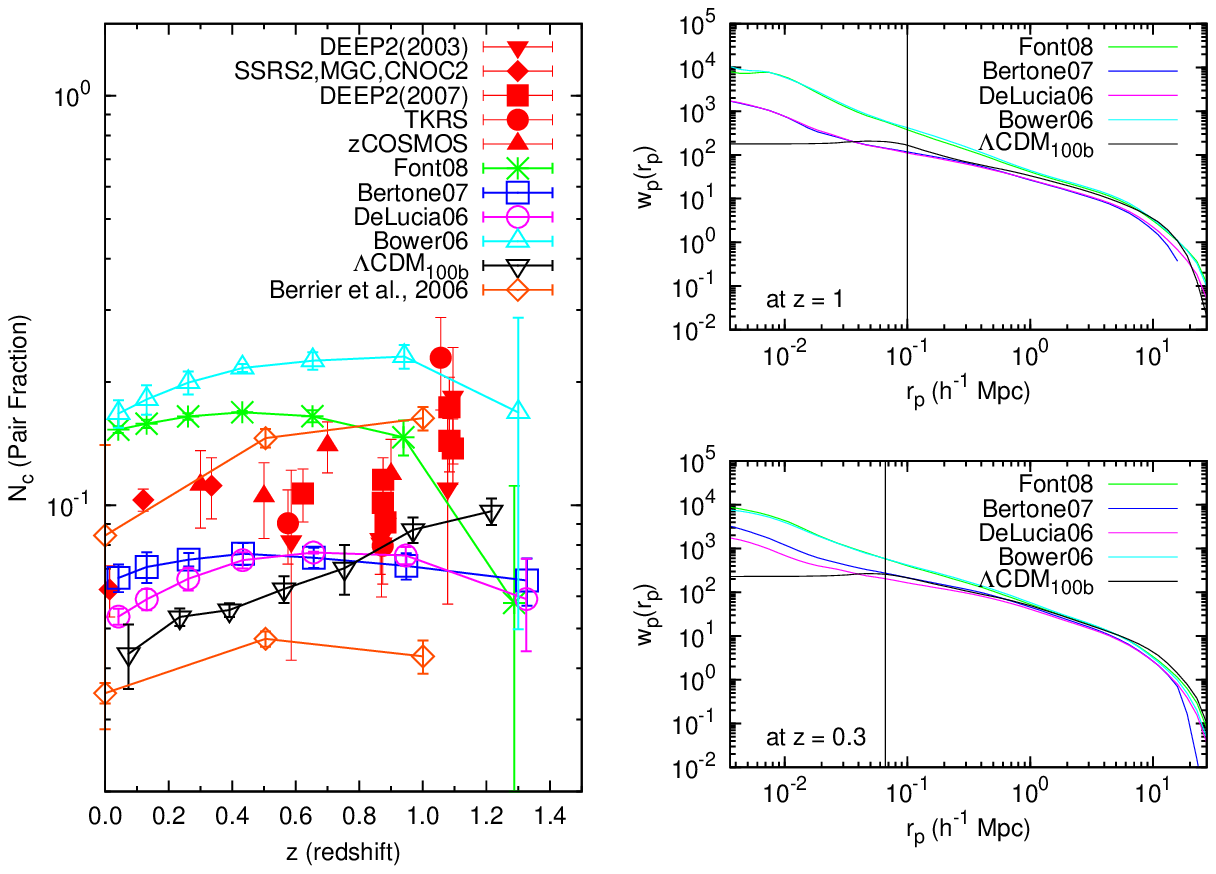}\\
  \caption{Left: The evolutions of the pair fraction $N_{c}$ computed with a evolved luminosity-selected criterion is shown. Observation results (red data points with errorbars) taken from \cite{lin04} include SSRS2 \citep{dac98}, CNOC2 \citep{yee00}, and DEEP2 early data \citep{dav03}, from full samples in \cite{lin08} include SRSS2, CNOC2, MGC \citep[Millennium Galaxy Catalog][]{lis03,dri05,all06}, TKRS \citep{wir04}, and DEEP2 \citep{dav03,dav07}, and from \cite{deRav11} (zCOSMOS). The values of the $N_{c}$ derived from the models are close to the observational data, and the discrepancy among the models is as larger as a factor of $\sim$ 3. However, their evolutionary trends appear to be flat in agreement with the observations. Right: Projected two-point correlation functions $w_{p}(r_{p})$ are plotted to show the clustering strength at small length scale where the vertical lines indicate the limiting physical projected separation 50 $\kpc$ for which pairs are counted. The observable abundance of close pairs can be estimated from $w_{p}(r_{p})$. Clearly, the stronger strength found at small length scale in Bower06 and Font08 reflects a higher value of $N_{c}$.
   }\label{fig:pf}
 \end{center}
\end{figure*}

\subsection{Pair fraction $N_{c}$ and galaxy merger rate}
There are two observational approaches used to probe the evolution of mergers. One is related to the close-pair count \citep[e.g. ][]{pat02,lin04,lin08}, and the other is to count galaxy mergers through morphological signatures of galaxy interaction \citep[e.g. ][]{con03,lotz08}. In this study, we follow the close-pair count approach to study the galaxy merger and explore its environmental dependence. In this approach, the direct observable is the average number of companions per galaxy, defined as
\begin{equation}\label{equ:pf}
  N_{c} \equiv \frac{2N_{pair}}{N_{g}},
\end{equation}
where $N_{pair}$ is the number of galaxy pairs and $N_{g}$ is the number of galaxies in the sample. Through studying $N_c$ from different models, a direct comparison between theories and observations can be made. Normally, the pair fraction is not equal to the average number of companions per galaxy, $N_c$, but in this study, we liberally refer the term "pair fraction" as "the average number of companions per galaxy". Following the observational criterion in \cite{pat02}, projected close pairs are so defined that the projected separation satisfies $10 \kpc$ $\leq$ $\Delta r$ $\leq$ $r_{max}$, where $r_{max}$ = 50 $\kpc$ is assumed in this study, and the rest-frame relative velocity $\Delta v$ is less than 500 $km ~ s^{-1}$. Additionally, galaxy samples are selected in the evolution-corrected B band magnitude range,$-21 \leq M^{e}_{B} \leq -19$, where $M^{e}_{B}$ is defined as $M_{B}(z=0) + Qz$, with $Q~=~1.3$ adopted from \cite{fab07}.

\begin{figure*}
 \begin{center}
  \includegraphics[width=15cm]{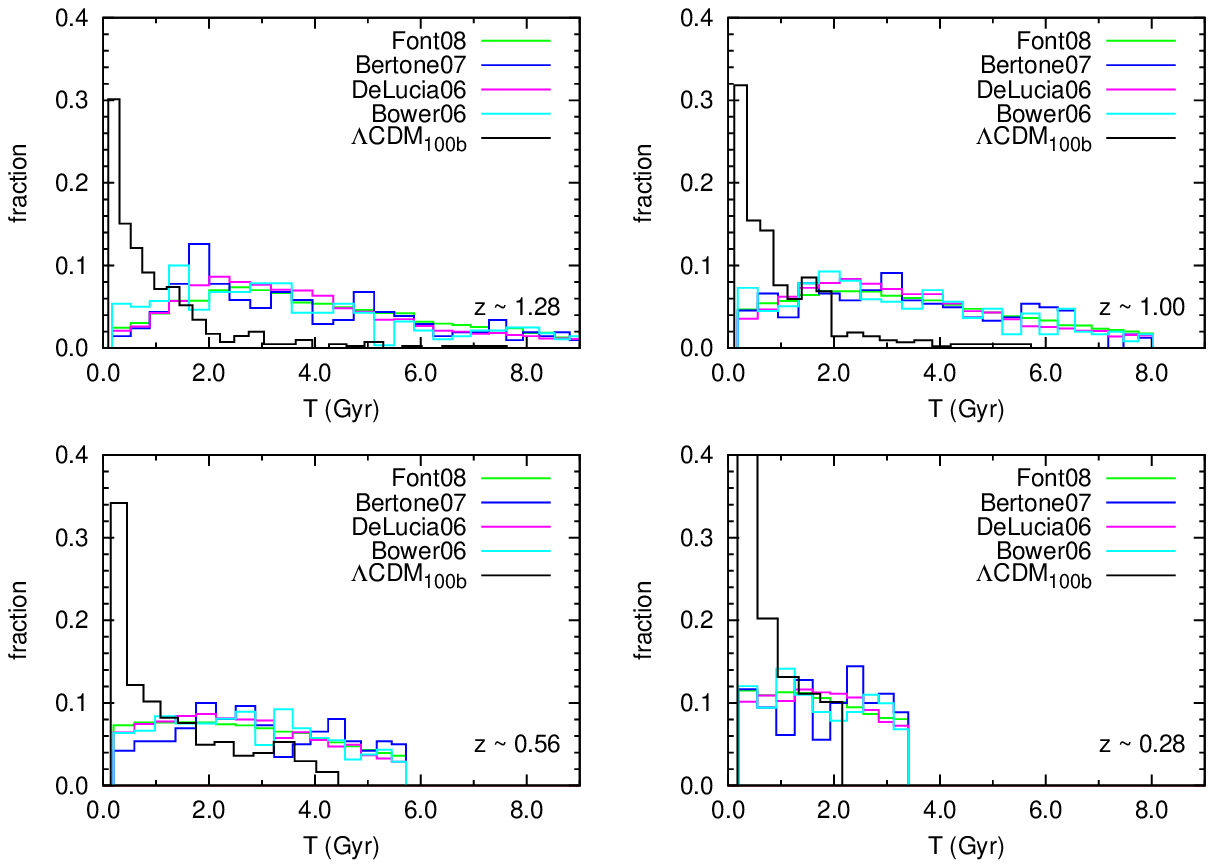}\\
  \caption{Distribution of the merging times in $Gyr$ at four different redshifts. Four SAMs are in good agreement. The distribution of the merging times gives a approximately constant profile from the SAMs, while there is a peak in the profile from $\Lambda CDM_{100b}$. Galaxy close pairs in $\Lambda CDM_{100b}$ have a large fraction of pairs that merge quickly leading to a peak at short merging time. The reason for the short merging time is mainly due to the lack of treatment on the orphan galaxies.
   }\label{fig:TmgDist}
 \end{center}
\end{figure*}

\begin{figure*}
 \begin{center}
  \includegraphics[width=15cm]{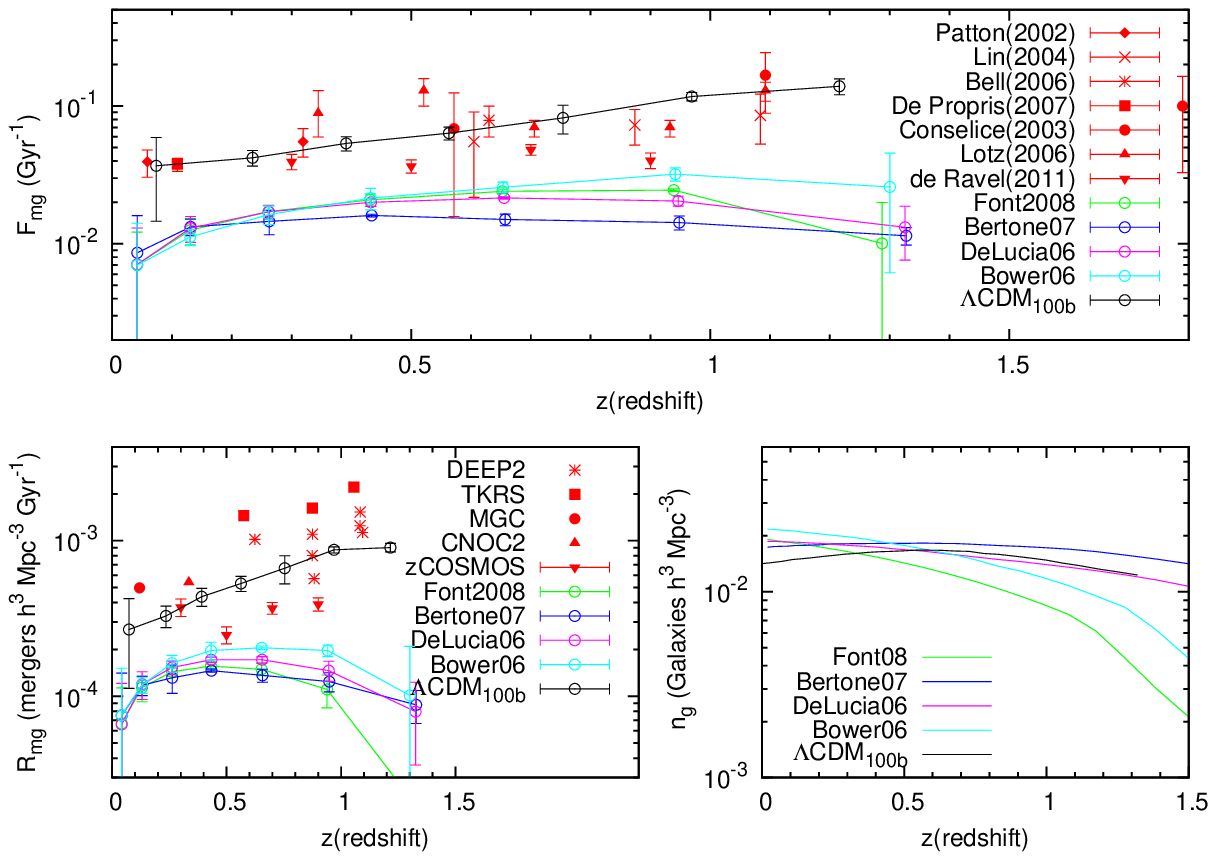}\\
  \caption{The normalized galaxy merger rate $F_{mg}$ (top), the volumetric merger rate $R_{mg}$ (bottom left), and the number density of galaxy sample in the magnitude range (bottom right) are plotted as a function of redshift $z$. Top: The observational results (red) are taken from literatures including those from \cite{pat02} (diamond), \cite{lin04} (cross), \cite{bell06} (star), \cite{dep07} (filled square), and \cite{deRav11} (filled reverse triangle) adopting the approach of galaxy pair counts, and from \cite{con03} (filled circle) and \cite{lotz08} (filled triangle) using a morphological analysis of galaxies. Bottom left: The observational data points are taken from literatures including \cite{lin08} (DEEP2, TKRS, MGC, and CNOC2) and \cite{deRav11} (zCOSMOS). The large deviation is observed between the observational merger rate and the theoretical values in the models. This is mainly due to the normalization factor $C_{mg}$/$T_{mg}$ adopted by the observations to convert $N_c$ into the merger rate much larger than those found in the models. However, the flat evolutionary trend is seen among the models. In addition, it is also evident that the merger rates vary with the models, and at high redshift the variation is as larger as an order of magnitude. Bottom-Right: The plot is to illustrate how the number density of galaxies varies with the redshift $z$. At high redshift, the deviation among different models is as large as a factor of ten. $R_{mg}$ basically is $F_{mg}$ $\times$ $n_g(z)$.
   }\label{fig:gmr}
 \end{center}
\end{figure*}

Our results on the evolution of the pair fraction $N_{c}$ are shown on the left panel in Figure~\ref{fig:pf}. This figure contains observation results (red errorbars) taken from \cite{lin04} and \cite{lin08} that include DEEP2 \citep{dav03,dav07} and some low redshift data, and simulation results from four semi-analytical models, Bower06, Font08, Delucia06, and Bertone07, along with results from \cite{jian08} ($\Lambda CDM_{100b}$) and \cite{ber06}. \cite{ber06} adopted two simple models for associating subhalos with galaxies: $V_{in}$, the maximum circular velocity that the subhalo had when it was first accreted into the host halo, and $V_{now}$, the maximum circular velocity that the subhalo has at the current epoch. Two models of \cite{ber06} give a reasonable range compared with observations, as shown in Figure~\ref{fig:pf}. It appears that the theoretical $N_{c}(z)$ vary with models by a factor as large as $\sim$ 3 and overall slightly deviate from the observations. However, the flat evolutionary trends in the models are consistent with the observations. The flat trend of $N_c(z)$ was first theoretically obtained by \cite{ber06} and our results from the models also reveal similar trends, suggesting that both observations and simulations support weak redshift dependence in $N_c$. Additionally, the projected close-pair counts are related to the projected two-point correlations, and hence stronger clustering implies more pair counts. To further examine the deviations found in $N_{c}$, the clustering of the galaxies at small scale satisfying the magnitude selection criterion from the models are explored. The projected two-point correlation functions $w_{p}(r)$ of different models are computed for comparisons and shown on the right panel in Figure~\ref{fig:pf}. It is seen that the clustering strength at small length scale varies with different models, and Bower06 and Font08 show stronger clustering than DeLucia06 and Bertone07 at length scale $r_{p}$ = 50 $\kpc$. The higher strength found at small length scale in Bower06 and Font08 indicates a higher value of $N_{c}$ in these two models.

\begin{figure*}
 \begin{center}
  \includegraphics[width=15cm]{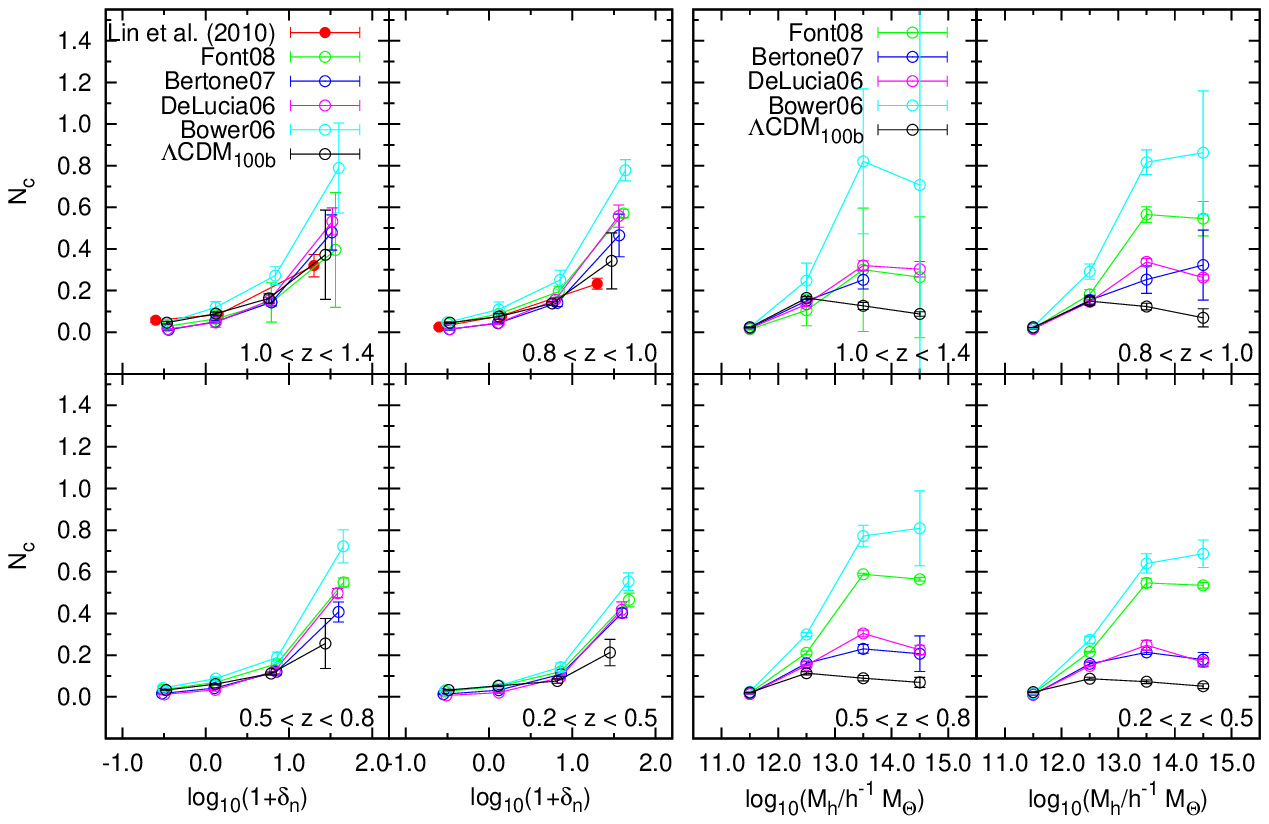}\\
  \caption{Pair fraction $N_c$ is expressed as a function of local density $(1+\delta_n)$ (left) and $M_h$ (right) in four redshift ranges. Because the correspondence between two estimators is not one to one, the profiles of $N_{c}$ from the models reveal the same linear growth with $(1+\delta_n)$ and $N_{c}$ is in good agreement with DEEP2 observation \cite{lin10}, while those of $N_{c}$ shows a turn-over in $M_{h}$. The result that higher $N_c$ found in high density region than in under-dense region is consistent with the expectation that galaxies are more clustered in dense region. It is also seen that $N_{c}$ varies with the models by a factor of $\sim$ 1.5 in high density region and by a factor of $\sim$ 4 in high $M_{h}$ bin.
   }\label{fig:pfDelta3}
 \end{center}
\end{figure*}

From observational point of view, galaxy merger rate is not an observable but an indirect quantity converted from the pair fraction $N_c$. On the contrary, the theoretical merger rate is a direct measurable in simulation. By taking this advantage, we connect the theoretical merger rate to the merger rate in observational form to derive $T_{mg}$ and to know the real meaning of $T_{mg}$ defined in observation. We start from giving a clear definition on how the merger rate is measured in this work. There are two definitions considered for the galaxy merger rate. The first one is the normalized merger rate, and it is defined and estimated as follows.
\begin{equation}\label{equ:mFmg}
  F_{mg} \equiv 2 \times \frac{\displaystyle\sum_{i=1}^{N_{mg}}\frac{1}{T_{i}}}{N_g},
\end{equation}
where $N_{mg}$ is the number of mergers, $T_i$ is the time to form the $i^{th}$ merger when the merger was still an identified close pair. In Equation~\ref{equ:mFmg}, the factor of 2 is introduced to turn the number of merger pairs into the number of merging galaxies. From Equation~\ref{equ:mFmg}, $F_{mg}$ shows dependence on the $T_i$ distribution of the merger pairs. To quantify the $T_i$ distribution, we follow the identified close pairs at four starting redshifts until they merge and record its distribution. The results are presented in Figure~\ref{fig:TmgDist}. The distributions among different SAMs are nearly flat and agree with each other. The flat distribution results in a longer medium or mean merging time about 3 to 4 $Gyrs$. On the contrary, the distribution from the simulation $\Lambda CDM_{100b}$ shows a peak at short merging time. The main reason for the deviation is correlated with the lack of orphan galaxies in $\Lambda CDM_{100b}$ and the merging time turns out to be much shorter than in the SAMs.

In contrast, the definition from \cite{lin10} for the normalized merger rate is
\begin{equation}\label{equ:fmg}
  f_{mg} = (1+G)\frac{C_{mg}N_{c}}{T_{mg}},
\end{equation}
where $G$ is the correction factor that accounts for the selection effect of companions due to the restricted luminosity range, and $C_{mg}$ represents the probability of galaxies in projected close pairs that will merge before the present. In mathematical expression,
\begin{equation}\label{equ:Cmg}
C_{mg} = N_{mg}/N_{pair}.
\end{equation}
  Because not all projected pairs are merging systems, $C_{mg}$ is introduced to account for the contamination from interlopers due to the difficulty in disentangling the Hubble expansion and the galaxy peculiar velocity. To obtain $f_{mg}$, $C_{mg}$ is hence as important as $T_{mg}$, and these two essential quantities are not direct observables but can be evaluated through simulations. Apart from the factor of $(1+G)$, $F_{mg}$ and $f_{mg}$ should be the same. The normalized merger rate $F_{mg}$ can then be expressed in terms of $N_c$, $C_{mg}$, and $T_{mg}$. Substituting $C_{mg}$ with Equation \ref{equ:Cmg} and $N_{c}$ with Equation \ref{equ:pf}, we then obtain that
\begin{equation}\label{equ:Fmgfmg}
 F_{mg} \equiv \frac{C_{mg}N_{c}}{T_{mg}} = \frac{2N_{mg}}{N_{g}T_{mg}}.
\end{equation}
Rewriting Equation ~\ref{equ:Fmgfmg}, $T_{mg}$ can then be expressed as a function of $F_{mg}$, $N_g$, and $N_{mg}$ such that
\begin{equation}\label{equ:Tmg}
    T_{mg}= \frac{2N_{mg}}{N_{g} F_{mg} } = \frac{N_{mg}}{\displaystyle\sum_{i=1}^{N_{mg}}\frac{1}{T_{i}}}=\frac{1}{\langle T^{-1} \rangle}.
\end{equation}
The determination of $T_{mg}$ is based on this formula in our work.

There is an innate difference between the true merger timescale, and the timescale over which the merger would be observable, i.e. the "observability timescale", see for example \cite{lotz11}. $T_{mg}$ is also an average observability timescale, but defined differently from Equation (8) in \cite{lotz11}. Our $T_{mg}$ definition is close to Equation (8) in \cite{kit08}, and the main difference is that they absorb $C_{mg}$ into the merger time but we do not, making their merger timescale longer than ours.

\begin{figure*}
 \begin{center}
  \includegraphics[width=15cm]{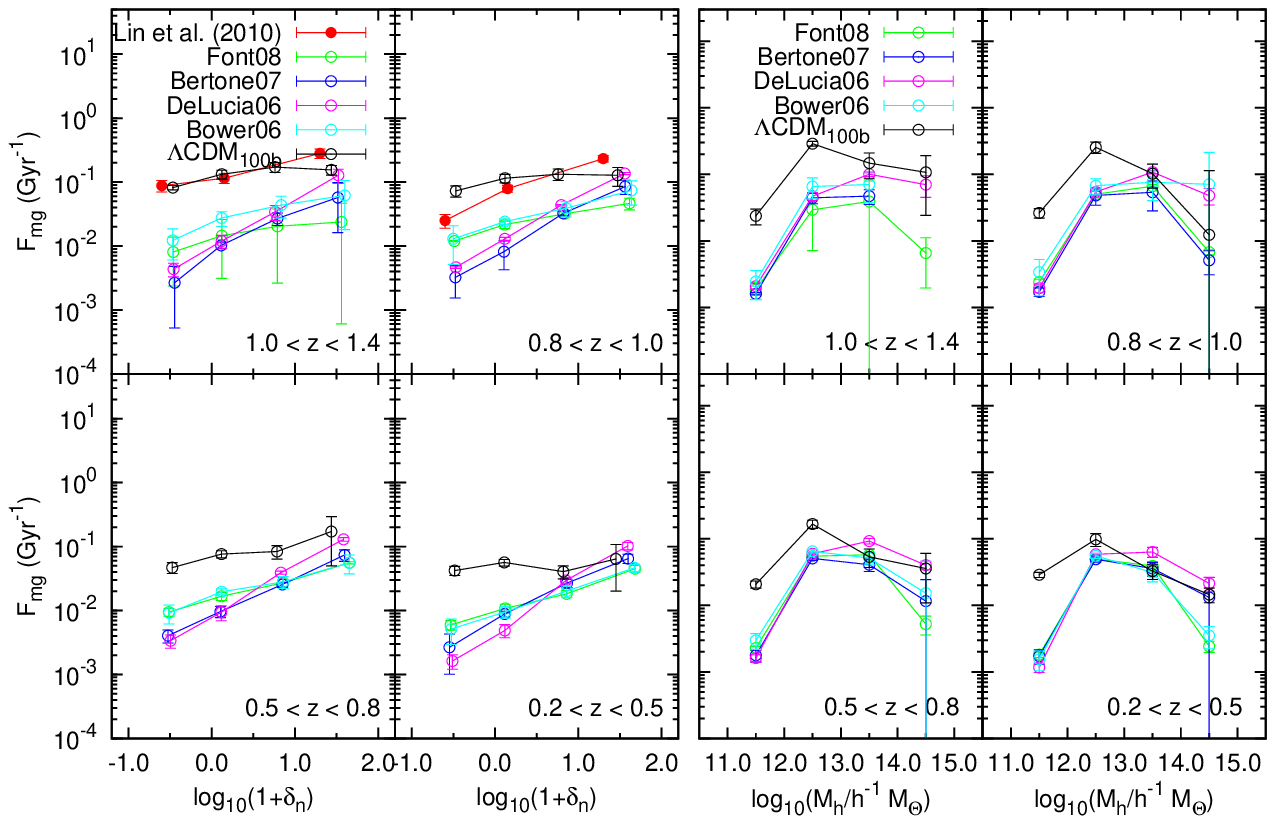}\\
  \caption{Normalized merger rate $F_{mg}$ is plotted as a function of local density $(1+\delta_n)$ (left) and $M_{h}$ (right). $F_{mg}$ reveals strong dependence on environments. Left: $F_{mg}$ increases with $(1+\delta_n)$, and in DeLucia06 (pink) and Bertone07 (blue) the increasing strength of $F_{mg}$ is over an order of magnitude in high density region than in low density region while in Bower06 (cyan) and Font08 (green), it is relatively flattened and is $\sim$ 4 time larger. Right: $F_{mg}$ reveals a turnover profile, and the difference between that of field galaxies and that of group galaxies is over an order of magnitude. Galaxies in group environment appear to have highest merger rate. The high merger rate in high local density regions is attributed to the significant contributions from galaxies in group environment.
   }\label{fig:FmgDelta3Mh}
 \end{center}
\end{figure*}

The other definition is the volumetric merger rate which has a form that
\begin{equation}\label{equ:mRmg}
   R_{mg}  \equiv \frac{ \displaystyle\sum_{i=1}^{N_{mg}}\frac{1}{T_{i}}}{L^3}= \frac{1}{2} F_{mg}  \times n_{g}(z),
\end{equation}
where $L$ is the length of the simulation box on a side in unit of $\Mpc$ and $n_g(z)$ is the comoving number density of galaxies, i.e. $n_g(z) = N_g/L^3$ in this study. Comparatively, in \cite{lin04} it gives that
\begin{equation}\label{equ:Rmg}
   \Gamma_{mg} = (0.5+G)\times \frac{n_{g}C_{mg}N_{c}}{T_{mg}},
\end{equation}
where $G$ is the correction factor previously described and the factor of 0.5 in $(0.5+G)$ converts the number of merging galaxies into the number of merger events. Except for the factor of $(0.5+G)$, $R_{mg}$ is the same as $\Gamma_{mg}$.

In Figure~\ref{fig:gmr}, the normalized merger rate $F_{mg}$ (top) and the volumetric merger rate $R_{mg}$ (bottom left) for the models and observations are plotted as a function of $z$, plus the evolution of the number density of the selected galaxies $n_g$ (bottom right). Deviations in the merger rates among the models are seen, but the flat evolutionary trends in the merger rate for the models are consistent in the figures of $F_{mg}$ and $R_{mg}$. With the parametrization $F_{mg}(z)~\propto~(1+z)^{m}$ in redshift range 0.2 $<$ $z$ $<$ 1.0, we find that $m$ $\sim 0.71 \pm 0.24$ for Font08, $\sim 1.88 \pm 0.19$ for Bower06, $\sim 0.359 \pm 0.31$ for DeLucia06, and $\sim 0.0 \pm 0.28$ for Betone07. When $R_{mg}$ is also assumed to be $\propto~(1+z)^{m}$, we find $m$ $\sim 0.46 \pm 0.38$ for Font08, $\sim 0.93 \pm 0.34$ for Bower06, $\sim 0.21 \pm 0.33$ for DeLucia06, and $\sim 0.0 \pm 0.34$ for Betone07. Besides, the large discrepancies between the models and observations are found mainly due to values of $C_{mg}$ and $T_{mg}$ adopted by observations. For example, \cite{lin04} adopts a normalization constant $C_{mg}/T_{mg}= 0.5/0.5$ = 1 while it is $\sim$ 0.15 from Font08 and Bower06, $\sim$ 0.2 from DeLucia06, and $\sim$ 0.17 from Bertone07. This also reflects a fact that the merger rate in the semi-analytical models implemented in Millennium simulation is comparatively smaller than those in other simulations. Moreover, the mass ratio of the merging galaxies has impact on the galaxy merger rate, with much higher merger rates when smaller galaxy mass ratios are considered, making comparison from observations to theory very uncertain. On the other hand, the result from the simulation $\Lambda CDM_{100b}$ appears to be broadly consistent with observational data points with slopes $m$ $\sim~2.18 \pm 0.11$ in $F_{mg}$ and $\sim~1.72 \pm 0.2$ in $R_{mg}$. The consistency arises primarily from the normalization constant $C_{mg}/T_{mg}$ found in $\Lambda CDM_{100b}$ to be close to that adopted in observations. The simulation $\Lambda CDM_{100b}$ uses a simplified galaxy model without considering the orphan galaxies that have already lost their surrounding subhalo, and thus it lacks contribution from these orphan galaxies when estimating the merger rate. It should be noted that the treatment for these orphan galaxies in semi-analytical models is such that once the subhalo disrupts, the orphan galaxy waits for a dynamical friction time before merging into the central galaxy of the main halo \citep[e.g. see][]{kit08}. That is, the merger timescale in the SAMs is much longer than that in $\Lambda CDM_{100b}$. Consequently, the merging history turns out to be very different, yielding in a higher merger rate in $\Lambda CDM_{100b}$. However, it was demonstrated by \cite{kit08} that with the orphan galaxy treatment, clustering at small scale at $z$ = 0 produces a similar correlation strength as in SDSS whereas there will be a deficit in the correlation strength without the treatment. The consistency of $\Lambda CDM_{100b}$ may therefore seem to be a coincidence and subhalo-subhalo mergers may present only a partial picture of real galaxy mergers. Our inclusion of the results from $\Lambda CDM_{100b}$ in this study is simply to provide a reference.

On the bottom-right panel in Figure~\ref{fig:gmr}, the evolution of $n_g$ with galaxies selected in two magnitude range is plotted for the models. The densities among the models can differ by an order of magnitude at high redshift and converge at low redshift. This plot is to demonstrate that under the same selection criterions, density variations exist among different models.

\subsection{$N_c$ and $F_{mg}$ as a function of environment}
In the previous section, redshift evolution of the pair fraction and the merger rate is discussed without considering environmental effect. To assess the effect of the environmental dependence of the pair fraction, $N_c$ is plotted as a function of local density $(1+\delta_n)$ (left) and $M_h$ (right) in four redshift ranges (a) $1.0~<~z~<~1.4$, (b) $0.8~<~z~<~1.0$, (c) $0.5~<~z~<~0.8$, and (d) $0.2~<~z~<~0.5$ in Figure~\ref{fig:pfDelta3}. When we bin galaxy close pairs into different environments, we exclude pairs whose pair galaxies are in different $(1+\delta_n)$ or $M_{h}$ bins, and we find that the fraction of the eliminated pairs in Font08 for $M_{h}$ bins, for example, is $\sim 15\%$ at $z \sim 1$ and drops to $\sim 6\%$ at $z = 0$. The fraction is small and unlikely to affect the result. On the left panel, it is seen that the pair fraction $N_c$ increases with the local density $(1+\delta_n)$ and different models are in good agreement with DEEP2 observation \citep{lin08}, except for the densest bin. The increasing tendency is understandable simply because galaxies in overdense environment are more clustered, and thus easily appear in pairs. By contrast, on the right panel $N_c$ appears to grow at low $M_h$ bins, but gradually gets flat and displays large deviations among the models at high $M_h$ bins. The explanation for this is that any given bin of $(1+\delta_n)$) is contributed from various $M_h$ bins, as discussed in Section~\ref{Mhdeltan} (see Figure~\ref{fig:MhDelta3}), and due to least contribution from the most massive cluster scale in overdense regions (the bins with high $(1+\delta_n)$), the total effect leads to a result that $N_c$ continues to grow as local density $(1+\delta_n)$ increases. Additionally, the evolution of $N_c$ appears to evolve weakly with redshift $z$ in any case. However, different SAMs differ considerably by a factor of $\sim$ 1.5 in $N_{c}$ in high density region and by a factor of $\sim$ 4 in the highest $M_{h}$ bin.

The normalized galaxy merger rate $F_{mg}$ is also evaluated in terms of $(1+\delta_n)$ (left) and $M_h$ (right) in Figure~\ref{fig:FmgDelta3Mh}. On the left panel, the observational results from \cite{lin10} are included for comparisons. It is evident that $F_{mg}$ has strong dependence on their surrounding environment, yielding that mergers occur more frequently in more dense region than in under-dense region. In addition, $F_{mg}$ appears to be nearly redshift independent but model dependent. On the left panel, if the galaxy merger rate is parameterized as $F_{mg}$ $\propto$ $(1+\delta_{n})^{\alpha}$, it is seen that $F_{mg}$ is flatter in Bower06 (cyan) and Font08 (green) with $\alpha~\sim~0.3$,  and steeper in DeLucia06 (pink) and Bertone07 (blue) with $\alpha~\sim~0.6~to~0.7$. That is, for all redshifts, galaxies merge more rapidly in high density regions than in under-dense regions by a factor of $\sim$ 4 in Bower06 and Font08 and by a factor of $\sim$ 20 in DeLucia06 and Bertone07. The positive correlation between galaxy merger rate and the locale density is in broad agreement with the measurement from \cite{fak09}, in that for galaxy-mass halos, mergers occur $\sim$ 2.5 times more frequently in the densest regions. The strong environment dependence is also consistent with the results in recent observational works \cite{lin10}, \cite{deRav11}, and \cite{kam11} over the redshift range 0.2 $<$ $z$ $<$ 1.2. In contrast, $F_{mg}$ in $\Lambda CDM_{100b}$ (black) shows nearly flat profiles implying that it is roughly independent of environment. This finding of no environmental dependence is close to what \cite{hes10} concluded using a merger tree of subhalos from the Millennium simulation. The difference between the SAMs and $\Lambda CDM_{100b}$ will be further discussed in Section \ref{discussion}.

\begin{figure*}
 \begin{center}
  \includegraphics[width=15cm]{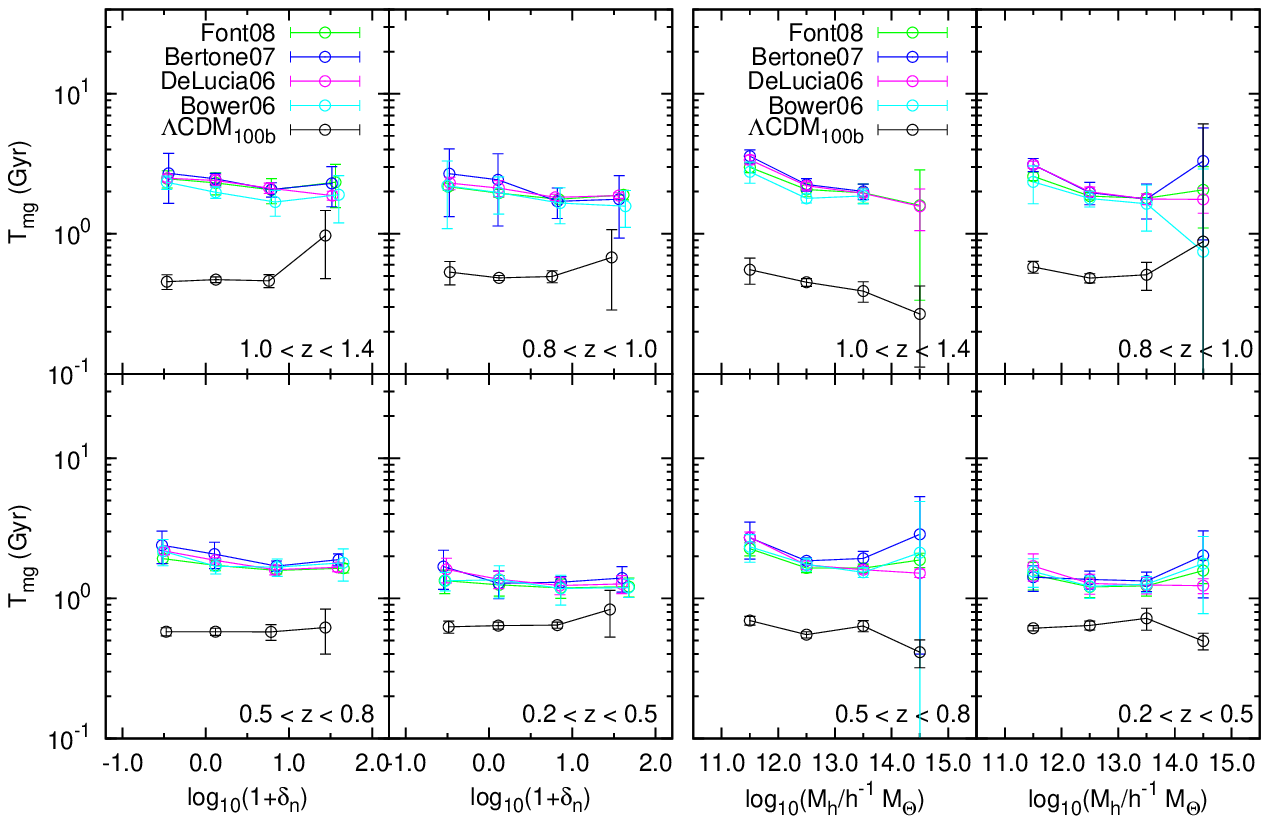}\\
  \caption{Merger time $T_{mg}$ as a function of local density $(1+\delta_n)$ (left) and $M_{h}$ (right). $T_{mg}$ is estimated using Equation~\ref{equ:Tmg}. Left: $T_{mg}$ is nearly constant without environment $(1+\delta_n)$ dependence, and agree well among the four SAMs. $T_{mg}$ in the high density region is 20$\%$ or less shorter than in the low density region. At lower redshift, $T_{mg}$ is apparently shorter. However, it doest not mean that projected close pairs found at low redshift merge more sooner, but $T_{mg}$ is computed from these merge before the present time and a short $T_{mg}$ will be obtained at low redshift. Right: $T_{mg}$ is also approximately independent of environment $M_{h}$ and the models. At high redshift, $T_{mg}$ in less massive $M_{h}$ bin (field environment) is twice longer than that in the most massive $M_{h}$ bin (cluster environment), but the errorbar in the bin is quite large. Therefore, $T_{mg}$ is assumed to be flat.
   }\label{fig:TmgDelta3Mh}
 \end{center}
\end{figure*}

On the other hand, when $F_{mg}$ is expressed in terms of $M_h$, the profile shows a turn-over and has a peak in the $M_h$ range between $10^{12}$ and $10^{13}$ $\Msun$, which corresponds to group environments. The major disparity among the models is at the most massive $M_h$ bin. Mergers occur approximately an order of magnitude more frequently in group or cluster environments than in field environment, and the peak indicates that galaxies in group environments merge most efficiently. Because galaxies in a local density $(1+\delta_n)$ bin are contributed partially from field, group, and cluster environments (see Figure~\ref{fig:MhDelta3}), the actual enhancement in the merger rate seen in high local density regions on the left panel of Figure~\ref{fig:FmgDelta3Mh} is mainly contributed from galaxies in group environment. This finding is in good agreement with that of \cite{tran08} that the group environment is critical for mergers to form massive galaxies. The fact that group galaxies have a high merger rate is expected, since in such environments galaxy close pairs have low velocity dispersion and are hence easier to merge. In addition, as we will discuss in Section \ref{CmgTmg}, close pairs in group environments are also less contaminated from the project effect.

\begin{figure*}
 \begin{center}
  \includegraphics[width=15cm]{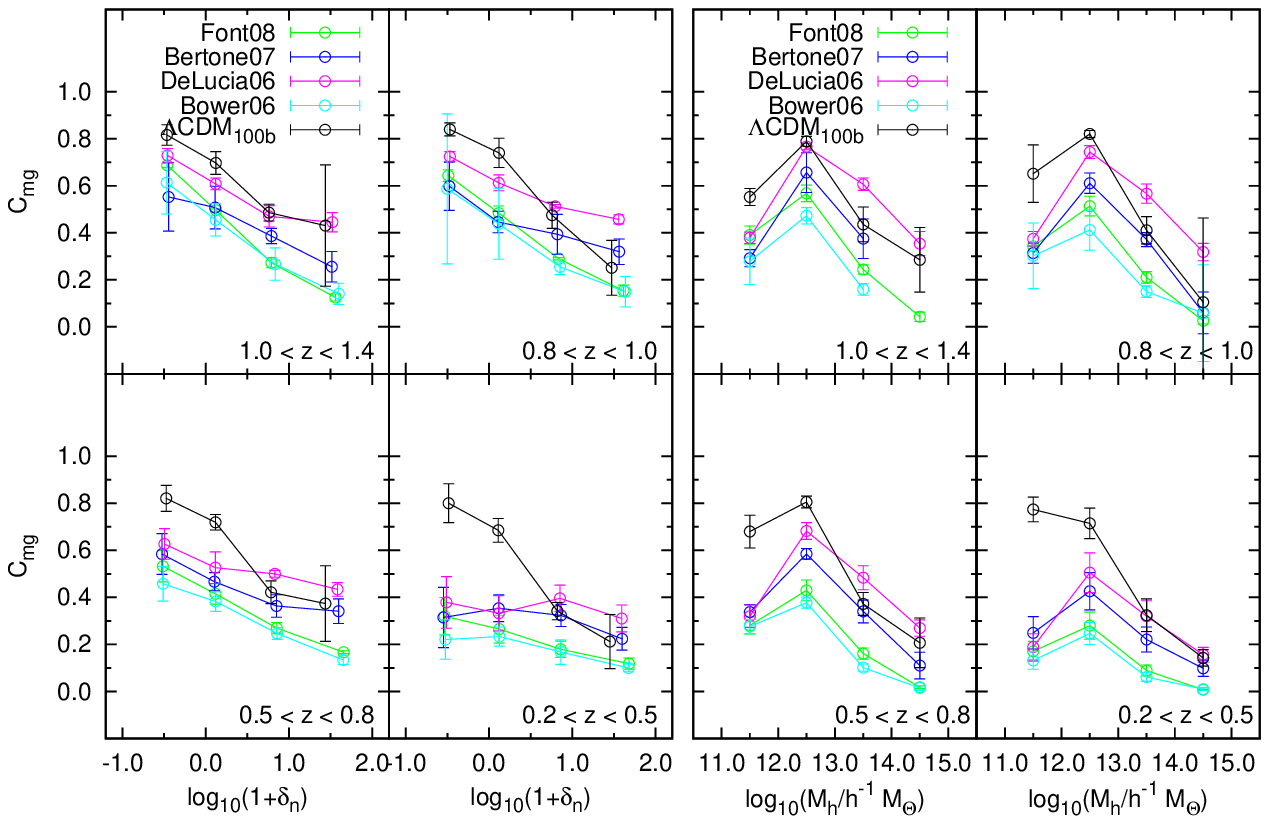}\\
  \caption{Merger fraction $C_{mg}$ as a function of local density $(1+\delta_n)$ (left) and $M_{h}$ (right). $C_{mg}$ is not only affected by the density, but also by the models. Left: $C_{mg}$ declines with the density. The variation between the lowest density and the highest density region is able to reach a factor of $\sim$ 5 in Bower06 and Font08, but a factor of $\sim$ 1.5 in DeLucia06 and Bertone07. At the same density bin, different model deviates as large as a factor of $\sim$ 3.5. Right: The highest value of $C_{mg}$ is in the bin with 12.0 $<$ $log_{10}(M_{h}/M_{\odot})$ $<$ 13.0 (group environment). The deviation among different models is as large as a factor of $\sim$ 10.
   }\label{fig:CmgDelta3Mh}
 \end{center}
\end{figure*}

\subsection{Environment dependence of the merger timescale $T_{mg}$ and the merger probability of close pairs $C_{mg}$} \label{CmgTmg}
Because $C_{mg}$ and $T_{mg}$ are two important quantities for observations to convert $N_c$ into galaxy merger rate and can be determined only theoretically, we explore them in detail in this section. We evaluate $T_{mg}$ from Equation~\ref{equ:Tmg} previously defined. In Figure~\ref{fig:TmgDelta3Mh}, $T_{mg}$ is expressed in term of two environmental estimators, $(1+\delta_n)$ (left) and $M_h$ (right). On the both panels, the merger time $T_{mg}$ shows weak dependence on environments. In the densest region, $T_{mg}$ is $\sim~10\%$ shorter than that in under dense region, and the deviation of $T_{mg}$ among different models is small. Moreover, the merger timescale $T_{mg}$ is much longer in the semi-analytical models than in $\Lambda CDM_{100b}$. The treatment on orphan galaxies to add dynamical friction time in the semi-analytical models is responsible for the longer timescale. We also observe that $T_{mg}$ depends slightly on redshift $z$ and declines as $(1+z)^{-1}$ for SAMs. However, we argue that the redshift dependence may not be a real effect. The shorter $T_{mg}$ obtained at low redshift is simply because many of projected close pairs identified at low $z$ do have no time to merge before the present. That is, the shorter $T_{mg}$ represents only a small merger population.

Regarding the evaluation on $C_{mg}$, we follow the merger tree forward in time in the simulations to examine whether projected close pairs will merge before the present. If close pairs do not merger before the present time, they will not be counted as mergers. The merger probability of close pairs $C_{mg}$ is the fraction of the merger number to the close pair number and is to account for the effect of interlope contamination.

The merger probability of close pairs $C_{mg}$ is plotted as a function of local density $(1+\delta_n)$ (left) and $M_{h}$ (right) in Figure~\ref{fig:CmgDelta3Mh}. Contrary to weak dependence on environments in $T_{mg}$, $C_{mg}$ reveals strong dependence on environments. On the left panel, $C_{mg}$ monotonically declines with the local density $(1+\delta_n)$ while it has a peak at the mid $M_h$ bin (12 $<~log_{10}(M_{h}/M_{\odot})~< $ 13), corresponding to the group environment. The profiles of $C_{mg}$ in different SAMs also show significant discrepancies by a factor of 3 on both left and right panels. At low redshift, the $C_{mg}$ profiles become flatter; this is understandable since a number of projected close pairs do not have enough time to merger before the present, and $C_{mg}$ does not vary with environment richness sensitively, thereby yielding a flatter profile and have lesser difference between underdense and overdense regions. In contrast, $C_{mg}$ at low $z$ in $\Lambda CDM_{100b}$ still shows a steep profile and differs with those of SAMs by a factor of $\sim$ 2 at low density regions. It is mainly due to the short merging timescale $T_{mg}$ in $\Lambda CDM_{100b}$ (see Figure~\ref{fig:TmgDist}), and resulting in a narrow distribution of merging time, and hence there is little difference between high and low redshifts.

\begin{figure*}
 \begin{center}
  \includegraphics[width=15cm]{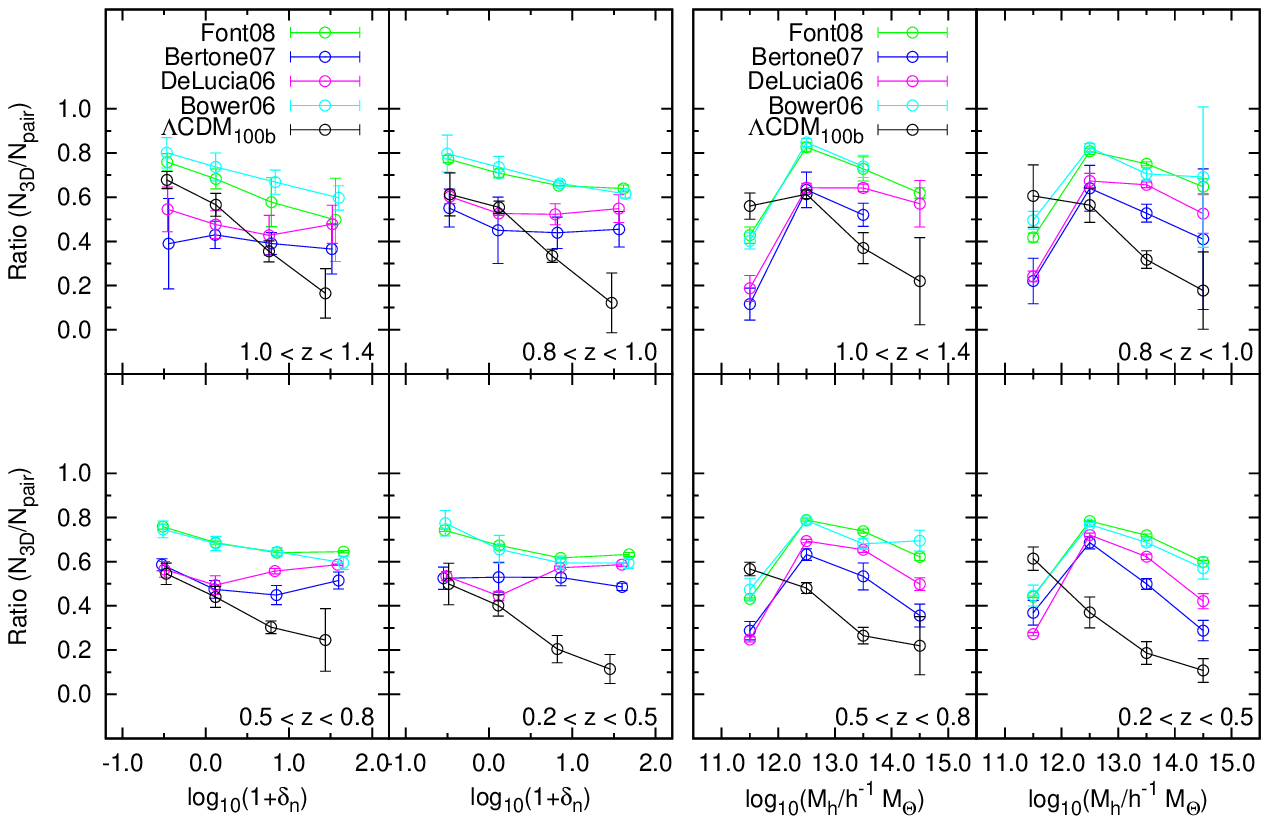}\\
  \caption{Ratio of $N_{3D}/N_{pair}$ as a function of local density $(1+\delta_n)$ (left) and $M_{h}$ (right). From this plot we are able to know in these 2D selected close pairs how many of them are three dimensionally close. It is found that the ratio has environmental dependence similar to $C_{mg}$, and it is thus inferred that the projection effect is responsible for the environmental dependence of $C_{mg}$. However, the slope of the ratio is flatter than that of $C_{mg}$, and it is suspected that some other physical mechanisms in dense environment prevent close pairs from merging.
   }\label{fig:projection}
 \end{center}
\end{figure*}

To understand the origin leading to the environmental dependence found in  $C_{mg}$, the ratio of $N_{3D}/N_{pair}$ is investigate to give an idea of how many projected close pairs are really close in three dimensional space, where $N_{3D}$ is the number of galaxy pairs that are close in three dimensional space in the 2D close pairs sample. The ratio is computed as a function of local density $(1+\delta_n)$ and $M_h$, and the results are shown in Figure~\ref{fig:projection}. The ratio of $N_{3D}/N_{pair}$ displays curves similar to the profiles of $C_{mg}$ on both right and left panels. It is thus inferred that the projection effect is very likely the main cause responsible for the drop of $C_{mg}$ in overdense region. However, the slope in the profile of the ratio of $N_{3D}/N_{pair}$ is not as steep as in that of $C_{mg}$. It is possible that in the overdense region projected close pairs are affected by some other physical processes, such as high velocity dispersion, to make merger harder to occur other than the projection effect.

\begin{figure}
 \begin{center}
  \includegraphics[width=12cm]{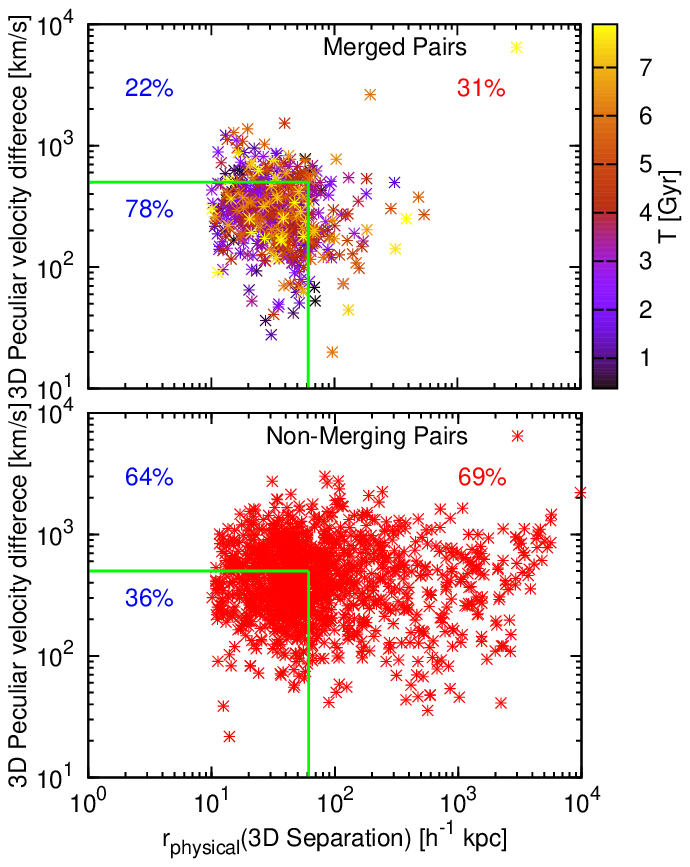}\\
  \caption{An illustration on how galaxies populate in the two dimensional parameters space of 3D physical separation and 3D velocity difference for those 2D selected close pairs that merge (top) and that do not merger before the present (bottom) at $z~\sim 1$ using Font08 as an example. On the top panel, the color bar indicates the merging time of the close pairs. Additionally, on top-right corner the red-color numbers $31\%$ and $69\%$ indicate the merger and non-merger probability of close pairs, respectively. The blue-color numbers inside and outside the box show the ratios that pairs satisfy and do not satisfy the projected close pair criterion. It is found that $\sim 64\%$ of the non-merger pairs are due to the projection effect, and $\sim 36\%$ of them are really 3D close pairs but fail to merge. The projection effect appears to be the main factor for close-pair contamination. It seems that it is likely high relative velocities to prevent 3D-close pairs from merging.
   }\label{fig:3Drvel}
 \end{center}
\end{figure}

To investigate this issue, we also plot three-dimensional separation versus three-dimensional velocity difference for those 2D selected close pairs that really merge (top) and that do not merge before the present (bottom) in Figure~\ref{fig:3Drvel} at $z~\sim 1.0$, using Font08 as an example to know how galaxies populate in this two dimensional parameter space. In this example, the merger and non-merger probabilities of close pairs are $\sim 31\%$ and $69 \%$, respectively. We find that for those merged pairs ($\sim 31\%$), about $78\% $ of them satisfy the close-pair selection criterions that the physical separation is less than $\sqrt{3/2}~50~\kpc$ and the velocity difference is less than 500 $km/s$, and $22\%$ of them are either with larger separation or large velocity difference but still merge in the end. On the other hand, for these non-merger pairs ($\sim 69\%$), about $36\% $ of them satisfy the close-pair selection criterions but do not merger in the end, and $64\% $ of them that are not three-dimensionally close and do not merge which can be attributed to the projection effect. The results suggest that the projection cause is the main cause for the close-pair contamination. However, surprisingly there are $25\%(=69\% \times 36\%)$ 3D-close pairs do not merge, suggesting that it is indeed high relative velocities between the 3D-close pairs to prevent them from merging.

\subsection{the impact of mass ratio of merging galaxies on the merger rate in different environments}

\begin{figure}
 \begin{center}
  \includegraphics[width=10cm]{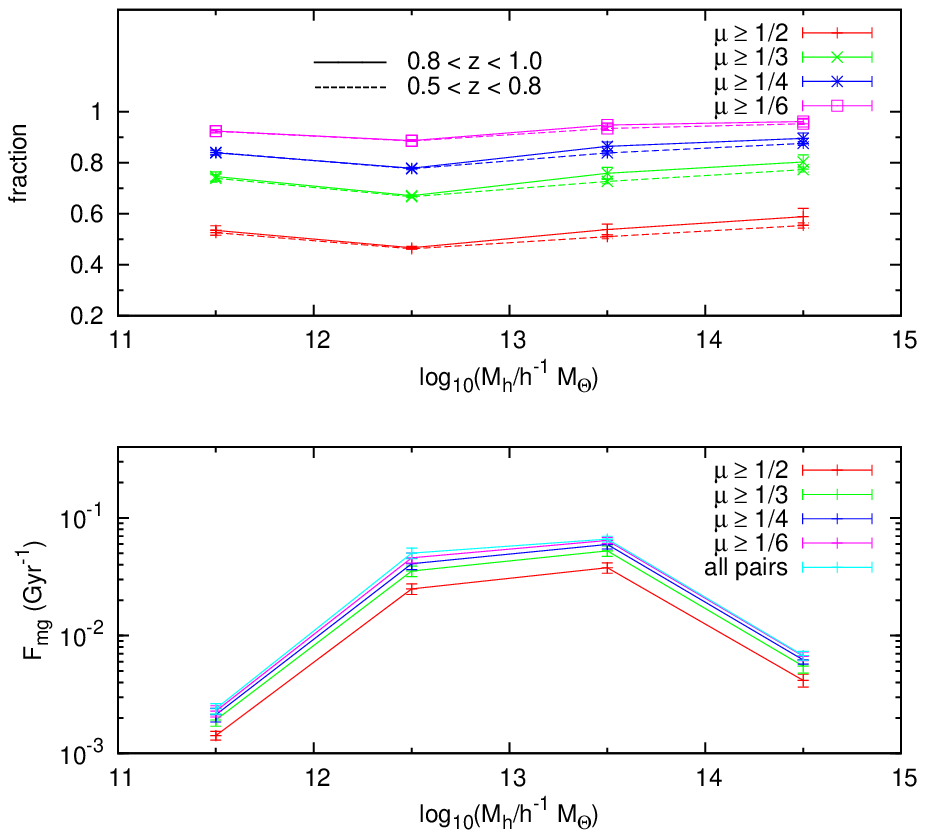}\\
  \caption{Top: The fraction of pairs with mass ratio $\mu$ $\geq$ 1/2 (red), 1/3 (green), 1/4 (blue), and 1/6 (pink) to total pairs as a function of $M_{h}$ at two redshift range, 0.5 $<$ $z$ $<$ 0.8 and 0.8 $<$ $z$ $<$ 1.0. The fraction shows roughly no environmental dependence implying that the $\mu$ composition in pairs is roughly the same no matter what environment they reside. Bottom: The galaxy merger rate $F_{mg}$ as a function of $M_{h}$ for different mass ratio criterions at redshift 0.8 $<$ $z$ $<$ 1.0. The similar shape among $F_{mg}$ profiles again infers that the impact due to mass ratio $\mu$ is not significant.
   }\label{fig:mrinmh}
 \end{center}
\end{figure}

There have been many discussions of merger rates and merger mass ratios \citep[e.g.][]{fak08,ste09,hop10}. To understand the impact of mass ratio of merging galaxies on the merger rate in different environments due to the luminosity cut sample selection in our study, we first estimate the fraction of pair satisfying the mass ratio criterion to total pairs as a function of environment in our luminosity cut, and we also examine its dependence on the mass ratio criterion. When we refer to merger mass ratios for pairs, we use the definition that $\mu$ = ($M_g^2$/$M_g^1$), the stellar mass of galaxy 2 over the stellar mass of galaxy 1, always taken such that 0 $<$ $\mu$ $<$ 1. In the top panel of Figure~\ref{fig:mrinmh}, the fraction of merger mass ratios satisfying $\mu$ $\geq$ 1/2, 1/3, 1/4, and 1/6 to all merger pairs is plotted as a function of $M_h$. The fraction exhibits roughly no environment dependence for the four merger mass ratios, implying that there is no significant discrepancy for the distribution of $\mu$ across different environments. Moreover, the fraction also shows no redshift evolution, suggesting that the passive evolutionary luminosity cut should select pairs of different epochs with similar physical properties. In addition, the galaxy merger rate $F_{mg}$ as a function of $M_{h}$ for different merger mass ratios is plotted in the bottom panel of Figure~\ref{fig:mrinmh}. We find that $F_{mg}$ increases as $\mu$ decreases, but the shape of $F_{mg}$ does not change. The similarity among different $F_{mg}$ profiles again indicates that the impact of mass ratio on the merger rate in different environments should be a minor effect.

\section{Discussion}
\label{discussion}

By using galaxy catalogs based on four SAMs, we find that the normalized galaxy merger rate $F_{mg}$ reveal strong environmental dependence and such dependence is similar at all redshifts but strongly model dependent. We observe that when the galaxy merger rate is expressed in terms of $(1+\delta_n)$ as $F_{mg}$ $\propto$ $(1+\delta_n)^{\alpha}$, $\alpha~\sim~0.3$ in Bower06 and Font08, and  $\alpha~\sim~0.6~-~0.7$ in DeLucia06 and Bertone07. The merger rate $F_{mg}$ between low density and high density regions can differ by a factor of $\sim$ 4 in Bower06 and Font08, and a factor of $\sim$ 20 in DeLucia06 and Bertone07. In contrast, $F_{mg}$ in $\Lambda CDM_{100b}$, representing a case of subhalo-subhalo mergers, shows less or no environmental dependence. Despite of large discrepancy in environmental dependence of $F_{mg}$ among the SAMs, the increasing tend is broadly consistent with the result found by \cite{fak09} that for galaxy-mass halos, mergers occur $\sim$ 2.5 times more frequently in the densest regions, whereas the result in $\Lambda CDM_{100b}$ is similar to what \cite{hes10} obtained that the merger rate does not correlate with environment. However, it may not be fair to make directly comparisons between our results and that of \cite{fak09} simply because our sample contains different halo populations whereas they did not. The fact that the positive correlation between $F_{mg}$ and the local density is observed in the SAMs and no correlation in $\Lambda CDM_{100b}$ arises mainly from competition between increasing $N_{c}$ and decreasing $C_{mg}$ with the local density with $T_{mg}$ remaining roughly constant. The pair fraction $N_{c}$ in the SAMs has a steeper increasing slope than that in $\Lambda CDM_{100b}$, while the merger probability of close pairs $C_{mg}$ in the models has a flatter decreasing slope than that in $\Lambda CDM_{100b}$. (We note that the enhanced $N_{c}$ at high local density in the SAMs is attributed to the presence of orphan galaxies.) In the end, the combined effect exhibiting in $F_{mg}$ turns out to be a flat profile in $\Lambda CDM_{100b}$ and a strongly increasing profile in the SAMs. As \cite{kit08} pointed out, inclusion of orphan galaxies is crucial when calibrating the conversion from pair counts to merger rates.

Deviation in the $C_{mg}$ and $T_{mg}$ determination among different SAMs is another interesting topic to be discussed. We notice that $C_{mg}$ and $T_{mg}$ adopted by observations vary with the referenced simulations, and the four SAMs we analyze give smaller $C_{mg}$ and longer $T_{mg}$ than those taken by observations \citep[e.g.,][etc]{pat02,lin04,lin08}. It leads to significant deviation in the merger rate between the observation inference and the model predictions. In our analysis, $C_{mg}$ ranges from 0.27 to 0.55 depending on the models and $T_{mg}$ is about 2 $Gyr$ nearly no model dependence at $z~\sim~1$. However, $T_{mg}$ is assumed to be 0.5 $Gyr$ in the observation work in \cite{pat02}. In addition to the larger projection separation 50 $\kpc$ taken in our study to derive $T_{mg}$ as opposed to 20 or 30 $\kpc$ taken by observations, the merger timescale is actually longer in the SAMs compared with simulation results from what \cite{lotz08b,lotz10} obtained,  $~\sim $ 1 $Gyr$. Our merging time distribution is wide and nearly flat (see Figure~\ref{fig:TmgDist}), and its median is $\sim$ 3 $Gyr$ at $z~\sim~1$. This value may be considered to be a representative merger timescale. If the merger timescale is estimated following the definition in \cite{kit08} that $\langle T_{merge} \rangle$ = $T_{mg}/C_{mg}$, it can result in an even longer timescale $\sim$ 2 to 4 $Gyr$. Additionally, \cite{lotz08b} gave a morphological analysis from a large suite of N-body/hydrodynamical gas-rich disc galaxy merger simulations processed through realistic radiative transfer models. For the same projected separation and velocity selection as in this paper, the time scale they obtained ranges from $\sim$ 0.3 to 1.4 $Gyr$ and the median is $\sim$ 1 $Gyr$. \cite{lotz10} further explored the effect of gas fraction, and mergers with different gas ratio result in time scale $\sim$ 0.9 $Gyr$ for 1:1 baryonic mergers, and $\sim$ 1.2 $Gyr$ for 3:1 baryonic mass ratio mergers. Therefore, depending on what simulation is adopted and what the definition for the timescale is, the uncertainty of merger rate can be as large as $\sim$ 4.

In addition to the uncertainty of the merger time $T_{mg}$ as discussed above, the merger probability of close pairs $C_{mg}$ also shows large variation depending on how they are calculated. Especially, the environmental dependence of $C_{mg}$ is addressed to a much lesser extent in the past. In \cite{lin10}, the decrease of $C_{mg}$ in high and low local density regions is by a factor of $\sim$ 4, approximately independent of $z$. However, this result corresponds to the case of subhalo-subhalo mergers. From our analysis, $C_{mg}$ is apparently model and redshift dependent. The decline is by a factor of $\sim$ 4 in Bower06 and Font08 and $\sim$ 1.5 in DeLucia06 and Bertone07 at redshift $z$ $\sim$ 1 while the decline is by a factor of $\sim$ 2 in Bower06 and Font08 and nearly flat in DeLucia06 and Bertone07 at $z~\sim~0.4$ . The uncertainty thus introduced by $C_{mg}$ could be as large as $\sim$ 2 to 3.

\section{Summary}
We make use of publicly available galaxy catalogs from four SAMs, including Bower06, DeLucia06, Bertone07, and Font08, implemented in the Millennium Simulation to explore the relation between the galaxy merger rate and its underlying environment, as well as to evaluate the model dependence of the merger rate. The approach taken by us closely follows the close pair observations by giving selections with an evolution-corrected B band magnitude range, $-21 \leq M^{e}_{B} \leq -19$, with the projection separation in physical length $\Delta r$ between 10 and 50 $\kpc$, and with the line-of-sight velocity difference less than 500 km/s. In this study, environment is quantified using the local density $(1+\delta_n)$ and the host halo mass $M_h$. Given these two estimators, we have been able to investigate the impact of environment on the galaxy merger rate. The results are summarized as follows.

(1) The whole $(1+\delta_n)$ distribution is decomposed into four bins based on the host halo mass $M_h$ that galaxies reside in to understand how these two environment estimators are correlated and how the correlation evolves with redshift. It is found that galaxies in more massive host halos tend to be have higher values of $(1+\delta_n)$. However, total number of galaxies in massive clusters is smaller than that in groups or small clusters. In other words, high density $(1+\delta_n)$ is dominated by galaxies from groups or small clusters.

(2) Although the pair fraction $N_c$ among the SAMs displays large discrepancies among them and slightly deviate from the observational data, their redshift evolution is consistently flat and in agreement with the same trend as the observational results and as the theoretical results of \cite{ber06}.

(3) The normalized merger rate $F_{mg}$ and the volumetric merger rate $R_{mg}$ are both below the observational results. This is mainly because the ratio of the merger probability of close pairs $C_{mg}$ to the merger time scales $T_{mg}$ (the normalization constant) found in the SAMs is much smaller than those adopted in observations.

(4) When the pair fraction $N_c$ is expressed in terms of environment, $N_c$ is found to be higher in higher density or more massive $M_h$ environment and the environmental dependence of $N_c$ evolves little with time. In addition, difference among different SAMs is found mainly in high density bins.

(5) We observe strong environmental dependence of the galaxy merger rate in the SAMs, where higher density regions have larger merger rates. When the galaxy merger rate is expressed as $F_{mg}$ $\propto$ $(1+\delta_{n})^{\alpha}$, the logarithmic slope is flatter in Bower06 and Font08 with $\alpha~\sim~ 0.3$ and steeper in DeLucia06 and Bertone07 with $\alpha~\sim~0.6~-~0.7$. That is, the merger rate between the low and high densities can differ by over an order of magnitude in certain SAMs. By contrast, the profile of merger rate as a function of halo mass is not power laws but exhibits more complicated dependence. The group environment turns out to be the regime where galaxies merge most frequently, consistent with what \cite{tran08} observed; high merger rate in high local density regions is, in fact, due to significant contributions from galaxies in group environments.

(6) The merger time $T_{mg}$ shows approximately no environmental and no model dependence. The difference between the densest and under-dense regions is $\sim$ $10\%$. Different environments have little influence on the merger timescale if these projected close pairs are true bound pairs. The merger time is found $\sim$ 2 $Gyr$ at z $\sim$ 1.

(7) Unlike $T_{mg}$, merger probability of close pairs $C_{mg}$ depends on environments and on SAMs. Despite the discrepancies among different SAMs, their profiles all have a similar trends that $C_{mg}$ drops as $(1+\delta_n)$ increase, and on the other hand, all SAMs have a peak in the bin with $log_{10}(M_{h}/M_{\odot})$ between 12 and 13.

(8) Through evaluating the ratio of real 3D close pairs to 2D close pairs, we find that the projection effect is responsible for the environmental dependence of $C_{mg}$. At $z~\sim~1$, only $\sim~31\%$ of projected close pairs will merge eventually by $z~=~0$. For those close pairs that do not merge, the projection effect appears to be the origin of major contamination.

(9) Because of no dependence on environments for the fraction of pair satisfying the mass ratio criterion to total pairs and the similarity among different $F_{mg}$ profiles, we conclude that the impact of mass ratio on the merger rate in different environments should be a minor effect.

\acknowledgments
H.-Y. Jian thanks A. Merson, J. Helly, C. M. Baugh, and R. Bower for helpful discussions on Millennium Simulation and SAMs during his visit in Durham. The work is supported in parts by the National Science Council of Taiwan under the grant NSC99-2811-M-002-138, NSC100-2811-M-002-128, NSC97-2628-M-002-008-MY3 and NSC99-2112-M-001-003-MY3.

%%%%%%%%%%%%%%%%%%%%%%%%%%%%%%%%%%%%%
% Tables
%%%%%%%%%%%%%%%%%%%%%%%%%%%%%%%%%%%%%

\end{document}